\documentclass[10pt,conference,letterpaper]{IEEEtran}
\IEEEoverridecommandlockouts
\usepackage{hyperref}
\usepackage{url}
\usepackage{cite}
\usepackage{amsmath,amssymb,amsfonts}
\usepackage{graphicx}
\usepackage{textcomp}
\usepackage{xcolor}
\usepackage{svg}
\svgsetup{inkscapelatex=false}
\usepackage{listings}
\usepackage{booktabs}
\usepackage{siunitx}
\usepackage{xcolor}
\usepackage{multicol}
\usepackage{lipsum}
\usepackage{subcaption}
\usepackage{algorithm}
\usepackage{algpseudocode}
\usepackage{amsmath}
\usepackage{multirow}
\usepackage{tabularx}

\definecolor{codegreen}{rgb}{0,0.6,0}
\definecolor{codegray}{rgb}{0.5,0.5,0.5}
\definecolor{codepurple}{rgb}{0.58,0,0.82}
\definecolor{backcolour}{rgb}{0.95,0.95,0.92}

\usepackage{caption}
\lstdefinestyle{mystyle}{
    backgroundcolor=\color{white},   
    commentstyle=\color{code},
    keywordstyle=[1]\bfseries\color{codegreen},%
    keywordstyle=[2]\bfseries\color{red},%
    numberstyle=\tiny\color{gray},
    stringstyle=\color{codepurple},
    basicstyle=\ttfamily\footnotesize,
    breakatwhitespace=false,         
    breaklines=true,                 
    captionpos=b,                    
    keepspaces=true,
    numbers=none,  
    frame=single,  
    frameround=tttt, 
    numbersep=5pt,                  
    showspaces=false,                
    showstringspaces=false,
    showtabs=false,                  
    tabsize=2
}

\lstdefinelanguage{OpenCL}%
{%
  morekeywords=[1]{__kernel,__global,__local,__constant,__private},%
  morekeywords=[2]{uint,uint2,uint3,uint4,uint8,uint16,int,int2,int3,int4,int8,int16,long,long2,long3,long4,long8,long16,float,float2,float3,float4,float8,float16,double,double2,double3,double4,double8,double16,size_t,uchar,uchar2,uchar3,uchar4,uchar8,uchar16,char,char2,char3,char4,char8,char16,void,bool,bool2,bool3,bool4,bool8,bool16,half,half2,half3,half4,half8,half16,short,short2,short3,short4,short8,short16,ushort,ushort2,ushort3,ushort4,ushort8,ushort16,discard,return,typedef,struct,switch,case,do,while,for,if,else,break,continue,goto,default,inline,static,extern,version,pragma,spirv,constant,sampler,sampler_t,event_t,error_t,status_t,clk_event_t,reserve_id_t,queue_t},%
  morecomment=[l]{//},%
  morecomment=[s]{/*}{*/},%
  morestring=[b]",%
  morestring=[b]',%
}%

\def\BibTeX{{\rm B\kern-.05em{\sc i\kern-.025em b}\kern-.08em
    T\kern-.1667em\lower.7ex\hbox{E}\kern-.125emX}}
\begin{document}

\title{LLMPerf: GPU Performance Modeling meets Large Language Models}

\author{\IEEEauthorblockN{Minh-Khoi Nguyen-Nhat\textsuperscript{*}}
\IEEEauthorblockA{\textit{FPT Software AI Center} \\
Hanoi, Vietnam \\
khoinnm1@fpt.com}
\and
\IEEEauthorblockN{Hoang Duy Nguyen Do}
\IEEEauthorblockA{\textit{FPT Software AI Center} \\
Hanoi, Vietnam \\
hoangdnd@fpt.com}
\and
\IEEEauthorblockN{Huyen Thao Le}
\IEEEauthorblockA{\textit{FPT Software AI Center} \\
Hanoi, Vietnam \\
huyenlt44@fpt.com}
\and
\IEEEauthorblockN{Thanh Tuan Dao}
\IEEEauthorblockA{\textit{FPT University} \\
Hanoi, Vietnam \\
tuandt28@fpt.edu.vn}
}

\maketitle

\begin{abstract}
Performance modeling, a pivotal domain in program cost analysis, currently relies on manually crafted models constrained by various program and hardware limitations, especially in the intricate landscape of GPGPU. Meanwhile, Large Language Models (LLMs) have demonstrated their effectiveness in addressing diverse programming challenges. Our work establishes a connection between LLMs and performance modeling, employing the LLM as a performance estimator. Through experimental exploration with carefully designed large-scale OpenCL datasets, we highlight the potential capability as well as the main difficulties of using LLMs in handling performance modeling tasks for OpenCL device source programs. As the first study for this line of work, our LLM-based performance model achieves a mean absolute percentage error of $24.25\%$ for a large-scale generated validation set. On a set of publicly available OpenCL programs, our model achieves a mean absolute percentage error of $46.1\%$.
\end{abstract}


\section{Introduction}
Performance modeling involves designing mathematical models to serve as evaluators or simulators for estimating the performance of applications. This task requires intensive analysis not only of the program characteristics themselves but also of the underlying systems and hardware architectures. The term "performance" can be understood in various ways, including program execution time, program throughput, or other relevant performance metrics. A GPU performance modeling task is a specific type of performance modeling focused on GPUs, which is a massively parallel environment. In this paper, we address the problem of predicting program execution time, specifically targeting OpenCL kernels due to their broad applicability across various machines, making the method more accessible for diverse hardware. The ability to accurately predict execution time is valuable for a wide range of tasks aimed at enhancing program performance, such as workload distribution and program optimization.

GPU performance modeling is generally categorized into two major approaches: analytical models \cite{hong2009analytical,li2015transit,wang2017cgpredict} and statistical models \cite{zhang2011performance,moolchandani2022performance}. CGPredict \cite{wang2017cgpredict} addresses the problem by leveraging the execution trace and memory trace of a single-threaded program to develop an analytical model specifically designed for the NVIDIA Kepler Architecture. These analytical approaches often rely on handcrafted models developed by domain experts, resulting in models with low error rates due to their careful design. However, these models require significant human effort and are prone to changes in system architecture \cite{madougou2016landscape}. To overcome these costly approaches, many statistical approaches developed. Overall, they aim to automate the model construction process by parameterizing the models and fitting their parameters to a training dataset collected from actual execution \cite{zhang2011performance}. 
Since previous statistical methods rely on traditional machine learning techniques, they operate on a limited scale of data and are sensitive to feature choices. Additionally, many statistical models require runtime information, either in the form of single-threaded runtime data \cite{karami2013statistical} or sample data points \cite{braun2020simple,arafa2021hybrid}, to achieve optimal performance. The reliance on dynamic features necessitates an available machine for kernel analysis and introduces runtime data collection overhead, thereby limiting the model's usability.

These limitations motivate the development of a statistical model that universally generalizes to any GPU program while considering only static information about the program and execution settings. In our setting, we fix the underlying system and only consider fundamental execution settings, including execution layout (global workgroup size, local workgroup size) and the kernel argument's value. Large Language Models (LLMs), recent advances in natural language processing, are well-suited for this purpose. LLMs, trained on extensive textual data, excel at understanding and generating human-like text and have demonstrated effectiveness in code-related tasks such as code auto-completion \cite{touvron2023llama} and bug fixing \cite{chen2023teaching}. By leveraging the code interpretability of LLMs and the awareness of interactions between code, hardware resources, and workload characteristics from a performance modeling perspective, our LLMPerf model aims to improve runtime predictions through more accurate reasoning about both code semantics and system-level performance factors.

Here, we formulate three questions that we think will be fundamental for future research:
\begin{enumerate}
    \item How do we make new AI models (such as LLMs) useful for performance modeling?
    \item How much can these models infer about the program without executing it?
    \item How do we effectively represent the program and its execution environment so that the new AI models can learn them?
\end{enumerate}

There are several key challenges to have accurate runtime estimation for arbitrary GPU source code. First, the massively parallel nature of the GPU programming architecture poses many difficulties in performance analysis. Second, the GPU architecture evolves quickly. This fact requires the performance models to attend to the low-level architectural changes and cover a wide range of performance factors. These performance factors are not always visible and require serious effort to analyze. Third, the GPU code has different performance behaviors with different inputs. It is not uncommon to see a library use different algorithms for the same problem with different input sizes \cite{adam2019pytorch}. Consequently, a dataset with high expressiveness is essential for model performance, requiring attention to not only program diversity but also input sizes and workgroup sizes. We address this by introducing a novel method for automatically generating a dataset that maintains expressiveness, leading to the creation of the first large-scale OpenCL performance dataset from real-life kernels. This dataset is then utilized to develop a pioneering LLM-based approach for estimating OpenCL execution times based solely on static information.

Our main contributions in this paper include:
\begin{enumerate}
    \item A framework for automated collection of the execution time of an arbitrary one-dimensional GPU kernel at any input size.
    \item A first large-scale OpenCL performance dataset from real life kernels
    \item A first large language model that estimates OpenCL execution time from only static information.
\end{enumerate}

Our paper is structured as follows: we describe our large-scale dataset creation framework for GPU kernel performance data in Section \ref{sec:dataset}. Section \ref{sec:model} introduces our proposed LLMPerf model that leverages LLMs for GPU kernel execution time prediction, followed by a comprehensive evaluation of LLMPerf's accuracy, efficiency, and generalization capabilities in Section \ref{sec:evaluation}. Finally, Section \ref{sec:conclusion} summarizes our key findings, limitations, and future research directions.

\section{A large-scale OpenCL Kernel Performance Dataset}
\label{sec:dataset}
Our dataset consists of two main components: a kernel source code corpus and the execution configuration for each kernel. The source code corpus, presented in section \ref{sec:source-corpus}, contains mainly kernel source code collected from various sources. The execution configuration, presented in Section \ref{sec:exe-config} contains the necessary information to execute the kernel on a target device. This information includes the global work size, the local work size, the input to the kernel function, and the corresponding host code of the kernel.
\subsection{Source Code Corpus}
\label{sec:source-corpus}
We adapt an OpenCL kernel corpus from BenchPress~\cite{tsimpourlas2022benchpress} with slight modifications in pre-processing: stripping double underscore prefixes and formatting using \verb|clang-format|. The corpus comprises over $15000$ OpenCL kernel source codes crawled from GitHub datasets.

The number of dimensions in a kernel's execution space is an important performance factor. Table~\ref{tab:dimensionality-distribution} shows the number of kernels belonging to each dimensional category in our corpus. Clearly, our corpus contains $10830$ one-dimensional kernels ($70\%$), accounting for the majority. This observation aligns with the intuition that most real-world problems are inherently one-dimensional, and higher-dimensional problems can be effectively reduced to a 1D representation. Consequently, we constrain our models to take 1D kernels as input, simplifying the learning scope while still capturing most real-life scenarios.

\begin{table}[htbp]
  \centering
  \caption{Dimensionality Distribution on all kernels in corpus}
  \label{tab:dimensionality-distribution}
  \begin{tabular}{|l|r|r|}
  \hline
  {Dimension} & { Count } & {Percentage (\%)} \\
  \hline
  {1D} & {10830} & {69.86} \\
  {2D} & {3879} & {25.02} \\
  {3D} & {793} & {5.12} \\
  \hline
  \end{tabular}
\end{table}

\subsection{Kernel Execution Configuration}
\label{sec:exe-config}
This section describes how we generate the kernel inputs and the kernel execution configurations so that the kernel source code can be compiled and executed in an automatic manner.

In reality, the performance is affected by the OpenCL program, the compiler, the operating system, and the host/device architectures.
Currently, we only consider the OpenCL programs and keep other factors fixed. Therefore, our models assume the performance variation only stems from the OpenCL programs.
Henceforth, we refer to the combination of these three elements \textit{(kernel source code, kernel inputs, and the kernel execution configuration)} as a \textbf{launch configuration}. This section describes our methods to generate the launch configurations in our dataset.

\subsubsection{Simple input argument selection}
\label{sec:simple-input-selection}
The kernel source lacks input argument information (sizes and values) passed from the host code. We generate this information, selecting input arguments based on the global work size \verb|gsize|:
\begin{enumerate}
\item If an argument is an array (e.g., \verb|__global float4*|), create an array of \verb|gsize| random values.
\item If an argument is a scalar (e.g., \verb|int2|), create a scalar equal to \verb|gsize|.
\end{enumerate}
We adopt \texttt{CLDrive}~\cite{cummins2017synthesizing} to implement this simple approach, generating valid input for $10250$ out of $15502$ kernels, assuming similar performance to the original input. It is important to clarify that a valid input in this context guarantees that the kernel can execute and produce its output without any compilation or runtime errors. Our underlying assumption is that the performance of the kernel with the generated input is similar to that of the kernel with the original input.

\subsubsection{Memory-analysis based input argument selection}
\label{sec:mem-anal-input-selection}

The simple approach can generate a large number of launch configurations. However, there are two main limitations. First, input array sizes and scalars are not always equal to the global work size. Second, scalar inputs can have varying values and affect performance, such as when they are part of branch divergence conditions. 

To address these limitations and generate kernels with wider performance characteristics, we propose an enhanced input argument selection strategy. Our approach analyzes the kernel source code to identify global memory access patterns and generate more diverse and realistic combinations of input array sizes, global work sizes, and local work sizes. From analyzing $100$ random kernels, we observe that most input array sizes can be inferred by statically analyzing memory access patterns, as GPU programs tend to have regular memory accesses to exploit parallelism. We identify four common patterns: data partitioning, offset and stride, boundary check, and complex forms combining the previous patterns.

\begin{itemize}
  \item \textbf{Data partitioning}: When the input size is large, it is common to partition the data into multiple chunks and assign each chunk to a work-item. The number of elements being processed by a work-item is often expressed by some constant \verb|k|. 
  A typical example of this pattern is the \verb|summation| program~\ref{lst:data-partition}, where a scalar \verb|k| specifies the number of elements on which each work-item should perform the summation operation.
\begin{lstlisting}[language=OpenCL, caption = Data partitioning, label = lst:data-partition]
__kernel void summation(__global float* in, __global float* out, int N, int k) {
    int id = get_global_id(0);
    float sum = 0.0f;
    for (int i = 0; i < k; i++) {
        int cId = id * k + i;
        if (cId < N) {
            sum += in[cId];
        }
    }
    out[id] = sum;
}
\end{lstlisting}

  \item \textbf{Offset and stride}: The memory indexes accessed in an array might not be consecutive, but rather interleaved with respect to the global ID, workgroup ID or local ID.
  For the most cases, the accessing index can be expressed in the form of two variables: offset and stride. 
  A common example of this pattern is the \verb|convolution| program~\ref{lst:offset-stride} that perform the convolution on a filter and a 1D-array. In this case, the offset is \verb|os| and the stride is \verb|st|.
\begin{lstlisting}[language=OpenCL, caption = Offset and stride, label=lst:offset-stride]
__kernel void convolution(__global float* in, __global float* out, int os, int st) {
    int id = get_global_id(0)*st+os;   
    out[id] = in[id - 1] * 0.5f 
                + in[id] * 1.0f 
                + in[id + 1] * 0.5f;
}
\end{lstlisting}

  \item \textbf{Boundary check}: 
  It is a common practice to have a boundary check in the kernel to guarantee no violated memory access. In this context, the input scalar value represents the size of the array that must be checked, preventing out-of-bounds errors and ensuring the integrity of the program's execution.
  Program~\ref{lst:boundcheck} shows a simple case where the kernel applies a boundary check for the array \verb|a| of size \verb|N|.

\begin{lstlisting}[language=OpenCL, caption = Boundary check, label=lst:boundcheck]
__kernel void initZero(__global float* a, int N) {
    int id = get_global_id(0);
    if (id < N) {
        a[id] = 0.0f;
    }
}
\end{lstlisting}
  
  \item \textbf{Complex form}: The last pattern is a combination of the three previous patterns. 
\end{itemize}

A common characteristic of the patterns is that each array index (hence the array size) is an affine function in the global ID, workgroup ID and local ID. 
We leverage this insight to propose an approach of fixing scalar values and finding a linear relationship between array size and global size. This allows decoupling the correlation between input size and global size, leading to more diverse input values for kernels.
Specifically, given a global size $gsize$ for kernel $K$ with scalar input argument list $S = \{s_1, \ldots, s_m\}$, where $s_i$  is the $i$-th scalar input, and $A = \{a_1, \ldots, a_n\}$, where $a_i$ is the $i$-th array input, we need to find:
\begin{equation}
size_{a_i} = c_{a_i} \cdot gsize + d_{a_i} \quad \forall a_i \in A
\end{equation}
considering fixed scalar values $\{value_{s_1}, \ldots, value_{s_m}\}$.

To find $(c_{a_i}, d_{a_i})$,
we sample two data points $(gsize^{(1)}, size_{a_i}^{(1)})$ and $(gsize^{(2)}, size_{a_i}^{(2)})$, which corresponds to the sizes of $a_i$ when running kernel $K$ with two sample global sizes $gsize^{(1)}$ and $gsize^{(2)}$. 
To obtain these points, we first use Clang AST to
instrument each memory access and record the memory access locations.
We then collect the memory access locations using two global work sizes $gsize^{(1)}$ and $gsize^{(2)}$.
We set the array sizes based on the smallest and largest recorded memory access locations. 
Obtained $(c_{a_i}, d_{a_i})$, we substitute $gsize$ into $size_{a_i} = ceil(c_{a_i} \cdot gsize + d_{a_i})$ to find the target array size $size_{a_i}$, where ceiling function $ceil$ takes a decimal number and rounds up it if its fractional part is greater than $0$. 

We choose input scalar values from a set of practically predefined values based on the observed patterns.
For instance, the boundary checks often set the scalars at a multiple of $gsize$; offsets and strides often use a small constant
such as $1$, $2$, $256$, \textit{etc.}. Therefore, we define a set of potential candidate values $cand_i$ for each $s_i$, and try all possible combinations in $cand_1 \times \ldots \times cand_m$. In our experiments, we set $cand_i = \{1, 4, gsize, 16, 32, 256\}$ for all $s_i \in S$. The algorithm \ref{algo:input-selection} details this process.
On average, a kernel has $2$ input scalar arguments. We find that this approach has a reasonable cost of running time.

\begin{algorithm}
\caption{OpenCL Kernel Input Arguments Selection Algorithm}
\begin{algorithmic}[1]
  \Require Kernel K, Global size gsize
  \State cands $\gets$ $\{1, 4, gsize, 16, 32, 256\}$
  \State scalars $\gets$ GetKernelScalarList($K$)
  \State arrays $\gets$ GetKernelArrayList($K$)
  \State K' $\gets$ InsertArrayHook($K$)
  \State combs $\gets$ $\text{cands}^{|\text{scalars}|}$
  \State validSettings $\gets$ $\{\}$
  \State exeSettings $\gets$ $\{(gsize1, lsize1), (gsize2, lsize2)\}$
  \ForAll{comb $\in$ combs}
    \State P1 $\gets$ RunKernel(K', combs, exeSettings[0])
    \State P2 $\gets$ RunKernel(K', combs, exeSettings[1])
    \State arraySizes $\gets$ $[0 \times |\text{arrays}|]$
    \ForAll{b $\in$ B}
    \State slope, intercept $\gets$ LinearInterpolation(P1[b], P2[b])
    \State arraySizes[b] $\gets \text{slope} \times \text{gsize} + \text{intercept}$
    \EndFor
    \If{all($\text{size} > 0$ for $\text{size}$ in $\text{arraySizes}$)}
    \State validCombs.append(\{comb, arraySizes\})
    \EndIf
  \EndFor
  \State \textbf{return} validCombs
\end{algorithmic}
\label{algo:input-selection}
\end{algorithm}

\subsubsection{Kernel Execution Settings Selection Strategy}
\label{sec:simple-execution-selection}
In this part, our goal is to find appropriate global work sizes and local work sizes so that the performance variation is most beneficial for the AI models to generalize the performance.

For the local work size (denoted as $lsize$), since a multiple of warp size to ensures that work-items in a work-group are distributed evenly into the warps. We enforce this constraint and randomize $lsize$ within the range of 
$[\texttt{warp\_size}, MAX\_LOCAL\_SIZE]$.

For each $lsize$, we vary $gsize$ to get the set of desired execution settings. 
Given a $lsize$ and considering all possible values of $gsize$, the corresponding performance of a GPU kernel can generally be categorized into three types: \textbf{idle SMs}, \textbf{under-utilized}, and \textbf{fully utilized}  sections~\cite{volkov2016understanding,dao2014performance}. 
Such categorization mainly stems from the ability of latency hiding in GPUs~\cite{volkov2016understanding}. 
Note that we can directly infer the $gsize$ value from the number of workgroups (denoted as $N_{WG}$) and $lsize$ using the following equation:
\begin{equation}
  N_{WG} = \dfrac{gsize}{lsize}
  \label{eq:wg-gsize}
\end{equation}
Furthermore, since running the kernels at the values of $gsize$ that are divisible by $lsize$ is more reasonable in reality, we use $N_{WG}$ to control the kernel parallelism for a given $lsize$ and derive $gsize$ from Equation~\ref{eq:wg-gsize}.

Let $N_{SM}$ denote the number of Streaming Multiprocessors (SMs) in the GPU. When $N_{WG} < N_{SM}$, not all SMs are active, leading to idle SMs during execution. The kernel execution time is typically the execution time of a single workgroup plus the overhead of workgroup launching. When $N_{WG}$ is very large (e.g., $N_{WG}>40 \times N_{SM}$), assuming round-robin SM scheduling, all SMs are busy during execution, and latency hiding is fully achieved in each SM. In this case, the kernel execution time is typically proportional to $N_{WG}$. When $N_{WG} > N_{SM}$ but is not sufficiently large to achieve fully utilized performance, the kernel is under-utilized. In this stage, all SMs are active, but the workload per SM is insufficient to hide all latencies. The GPU throughput increases as $N_{WG}$ grows. Among the three types, under-utilized execution configurations contain the most performance information due to varying parallelism per SM and characterizing the OpenCL workgroup scheduler. Therefore, we give the highest priority to sampling $N_{WG}$ for this type. Idle SMs generally contain more performance variation than fully-utilized, as fully utilized execution times linearly increase with workload. Based on these considerations, we use an empirical sampling ratio of $25:60:15$ for idle SMs, under-utilized, and fully-utilized types, respectively. To detect the fully utilized point, we use a common constant $k \times N_{SM}$ to avoid significant profiling overhead.

\textbf{Handling data imbalance}.

There are two main limitations with the aforementioned analysis: it does not consider the kernel execution time and the exact boundaries of the three performance types significantly depend on the kernel characteristics. 
Consequently, the generated data suffer from data imbalance with regard to the execution time distribution of the generated launch configurations. This is a non-trivial hurdle for the AI models to learn the performance from the samples.

\begin{figure}[hb]
  \centering
  \begin{subfigure}[b]{.5\linewidth}
    \includegraphics[width=\linewidth]{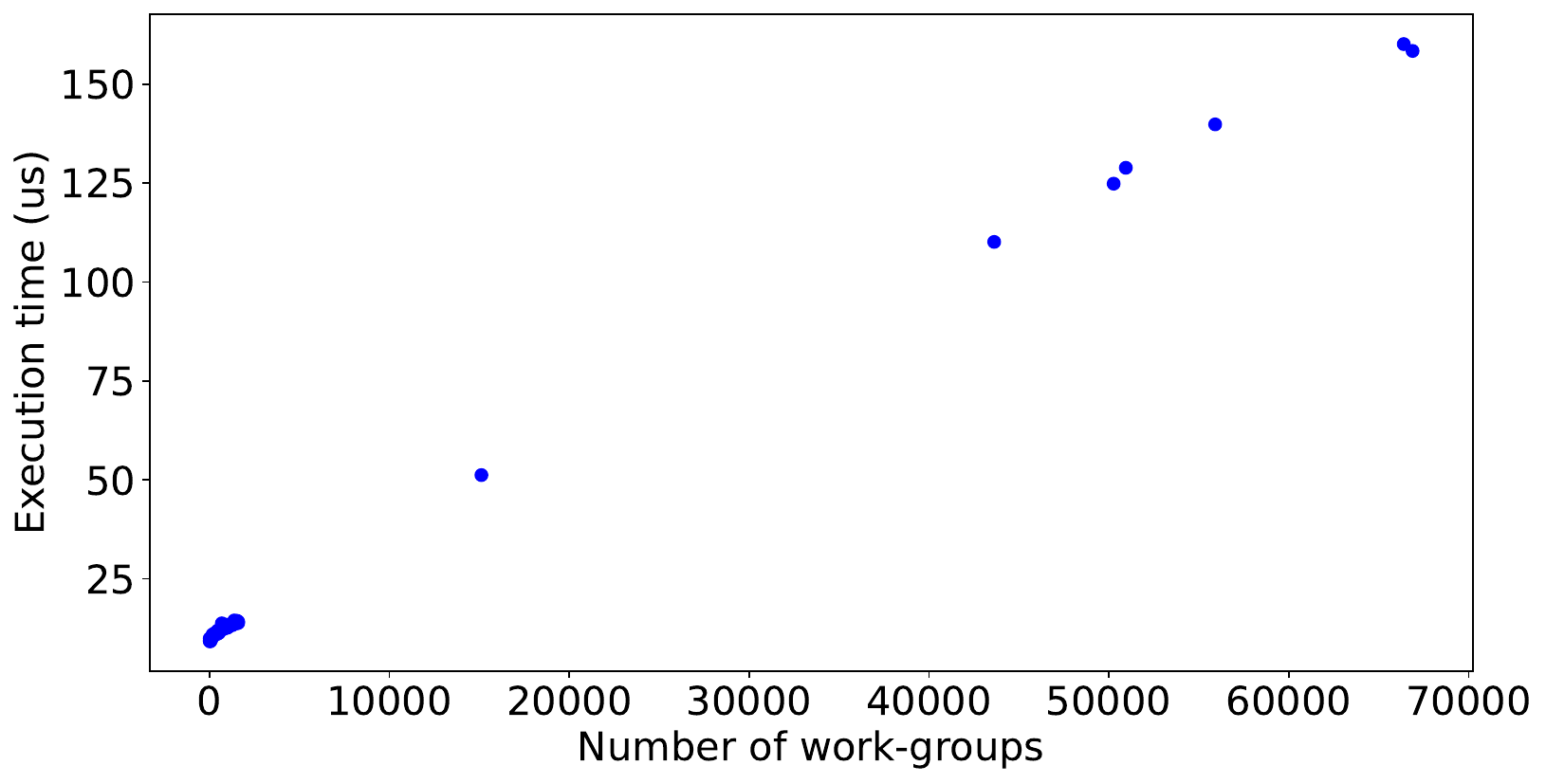}
    \subcaption{Before using IQR}
    \label{fig:data-refine-ker-small}
  \end{subfigure}%
  \begin{subfigure}[b]{.5\linewidth}
    \includegraphics[width=\linewidth]{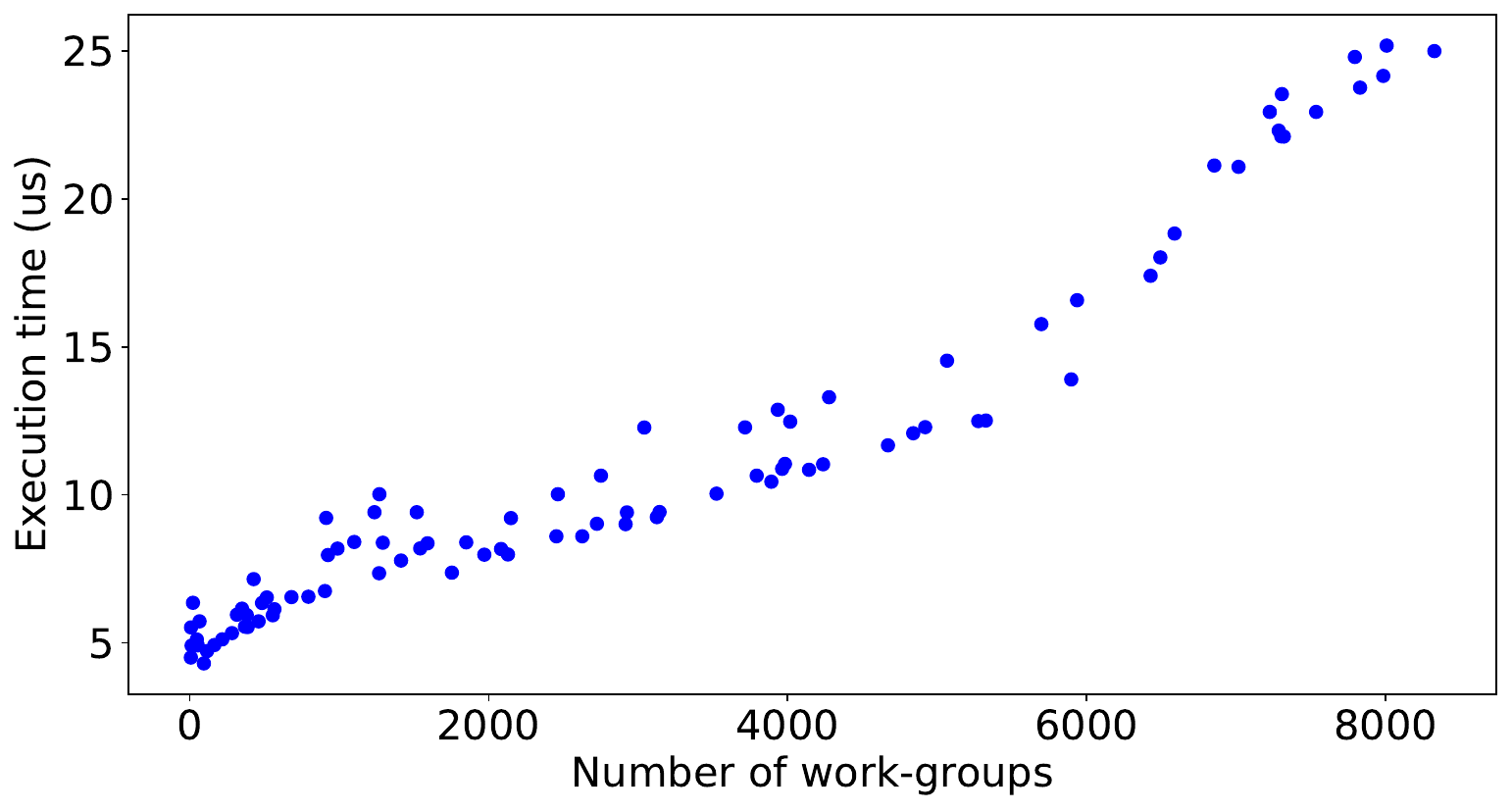}
    \subcaption{After using IQR}
    \label{fig:data-refine-ker-full}
  \end{subfigure}
  \caption{Example of execution time data before and after using IQR}
  \label{fig:data-refine}
\end{figure}

Figure~\ref{fig:data-refine-ker-small} shows the execution time distribution of kernel \textit{division\_assignment\_scalar} in our corpus. The graph shows the execution times with $lsize=128$ and varying $gsize$.
The majority of the data lies on the bottom-left corner of the graph, \textit{i.e.,}, with very small execution times, while only a few data points that have the medium and long execution times.

To handle the imbalance problem, we propose a refinement algorithm to generate more balanced data points with regard to the execution time. 
The idea is based on an outlier detection method called the Interquartile Range(IQR)~\cite{vinutha2018detection}. 
The algorithm assumes the data has a normal distribution. It estimates the central of the distribution based on frequent data points (small execution times) and the two edges of the distribution based on the infrequent data points (medium and long execution times). The algorithm then estimates the range of execution times missing from the distribution.
We fit a linear regression model on the data generated and use this regression model to derive the corresponding $N_{WG})$ to sample from the missing execution times.
We can observe the improvement of the execution time distribution in Figure \ref{fig:data-refine-ker-full}.

\section{LLMPerf Model}
\label{sec:model}
This section provides information of our model architecture, input representation, and training process that we use to train the model for execution time estimation. We emphasize that our model does not use any runtime information. It functions without any requirement of the OpenCL or GPU execution environment. In the inference phase, the only input it requires is the source code of the OpenCL kernel.

\subsection{Prompt Engineering}
Our model is a Large Language Model (LLM), thus we need to represent the input as a prompt, which is a text structure that the model will take as input. 
Designing an effective prompt is crucial, as the model learns better with a well-structured prompt. 
We leverage the prompt technique from the popular instruction-tuning dataset Alpaca~\cite{alpaca} with modifications to better fit our use case. 
Specifically, for one data sample, the prompt is formatted as follows:

\begin{lstlisting}
Predict time for the following OpenCL kernel:
{code}
Given input:
{inputs}
Given kernel are executed with global size equals cl::NDRange({global_size}), local work-group size equals cl::NDRange({local_size})
\end{lstlisting}

In the prompt above, \verb|code| section contains the kernel source code.
The \verb|input| section contains the input argument settings generated by concatenating the description of each input argument in the kernel on a separate line. 
We have separate description templates for array and scalar input arguments:
\begin{lstlisting}
Argument at position {id} is {name}, which is {qualifier} buffer of type {type} with size {value}
\end{lstlisting}
\begin{lstlisting}
Argument at position {id} is {name}, which is {qualifier} scalar of type {type} with value {value}
\end{lstlisting}

Here, \verb|id|, \verb|name|, \verb|qualifier|, \verb|type|, and \verb|value| provide the argument index, argument name, argument qualifier (\textit{e.g.}, \verb|global|, \verb|local|, \verb|const|, and \verb|private|), argument data type (\textit{e.g.}, \verb|int|, \verb|float|), and argument size (if the argument is an array) or argument value (if the argument is a scalar), respectively.

This prompt format captures the essential static information needed for our model to predict execution time, with a focus on array sizes rather than element values. This choice is made for three key reasons: (1) OpenCL programs typically process large datasets in a SIMD manner, where the amount of data is the dominant performance factor; (2) while array element values can impact performance, especially in cases of memory or execution divergences, such scenarios are generally avoided by OpenCL programmers; and (3) including entire array values would make the prompt excessively long, risking context length limits in LLMs and complicating the optimization process.

\subsection{Model architecture}
Large Language Models (LLMs) are pre-trained on vast textual data using self-supervised learning objectives and often require further fine-tuning on task-specific datasets to adapt their knowledge. We choose CodeGen~\cite{nijkamp2022codegen} as the pre-trained model and consider execution time estimation a fine-tuning task. CodeGen is a powerful LLM family developed by Salesforce Research for code-related tasks, trained on a vast corpus of source code and natural language data.

To better fit the regression task, we remove the language prediction head from CodeGen and introduce a new regression head to allow direct execution time prediction for a given kernel. This regression head takes $N$ vectors of dimension $d$, concatenates all $N$ vectors into one feature vector of dimension $Nd$, and projects the feature vector to the target execution time using a Multi-Perceptron (MLP) layer. We call this model LLMPerf.
LLMPerf is trained on the generated prompt dataset to predict the binary logarithm of execution time using Mean Squared Error (MSE) loss. We chose to predict the logarithm of execution time to reduce sensitivity to large execution times and stabilize the training process.

\section{Evaluation}
\label{sec:evaluation}
\begin{table}[htbp]
    \centering
    \caption{Datasets information with number of samples after preprocessed inside brackets}
    \label{tab:dataset-info}
    \begin{tabular}{|l|c|c|c|}
    \hline
    & \textbf{200K} & \textbf{230K} & \textbf{400K} \\
    \hline
    \textbf{Train} & 177892 (171790) & 206369 (198689) & 348970 (338434) \\
    \textbf{Val} & 19922 (19655) & 23500 (23233) & 39975 (38797) \\
    \hline
    \textbf{Total} & 197814 (191445) & 229869 (221922) & 388945 (377231) \\
    \hline
    \end{tabular}
\end{table}
In this section, we provide an evaluation of our LLMPerf model. We describe our performance datasets in Section~\ref{sec:evaluation-opencl-dataset}. Section~\ref{sec:evaluation-experiment-large-dataset} details the experimental setup and results on our large-scale dataset.

Section~\ref{sec:evaluation-experiment-public-benchmark} assesses the model on well-established OpenCL benchmarks, showcasing its ability to handle a wide range of kernels beyond the training data.
\subsection{OpenCL performance dataset}
\label{sec:evaluation-opencl-dataset}
In order to demonstrate the effectiveness of our data generation techniques, we create three versions of datasets.
We use NVIDIA Tesla V100S PCIe 32 GB ($N_{SM} = 80$) to generate the performance data.
We use the OpenCL C++ interface~\cite{cppopencl,cummins2017synthesizing} to create host programs and run kernels with OpenCL 2.2.
\begin{table}[htbp]
    \centering
    \caption{Experiment of LLMPerf on three large-scale datasets}
    \label{tab:eval-large-dataset}
    \begin{tabularx}{\linewidth}{|l|X|X|X|}
    \hline
    \textbf{Model} & \textbf{Train MAPE} & \textbf{Val MAPE} & \textbf{Val MAPE (400K)} \\ 
    \hline
    LLMPerf-350M-200K & 23.29 & 42.68 & 80.53 \\
    LLMPerf-350M-230K & 21.78 & 37.39 & 51.48 \\
    LLMPerf-350M-400K & 1.86 & 43.65 & 43.77 \\
    LLMPerf-2B-200K & 6.55 & 34.2 & 64.9 \\
    LLMPerf-2B-230K & 7.09 & 23.61 & 54.44 \\
    LLMPerf-2B-400K & 3.92 & 24.25 & \textbf{24.25} \\
    \hline
    \end{tabularx}
\end{table}

\textbf{200K}. This dataset contains 200K data points generated using the simple version of the input argument selection strategy.
We set $k=80$. For each kernel source code, we generate $200$ launch configurations ($4$ random $lsize$ and $50$ random $gsize$).
We generate one input set for each of $200$ launch configurations.

\textbf{230K}. We employ the memory analysis based input argument selection strategy to generate $30K$ additional data points. 

\textbf{400K}. Lastly, we apply IQR to the kernel execution setting selection of the $230K$ dataset to create a total of $400K$ data points with a more balanced data distribution.

\begin{figure*}[h]
    \centering
    \begin{subfigure}[b]{0.3\textwidth}
        \includegraphics[width=\linewidth]{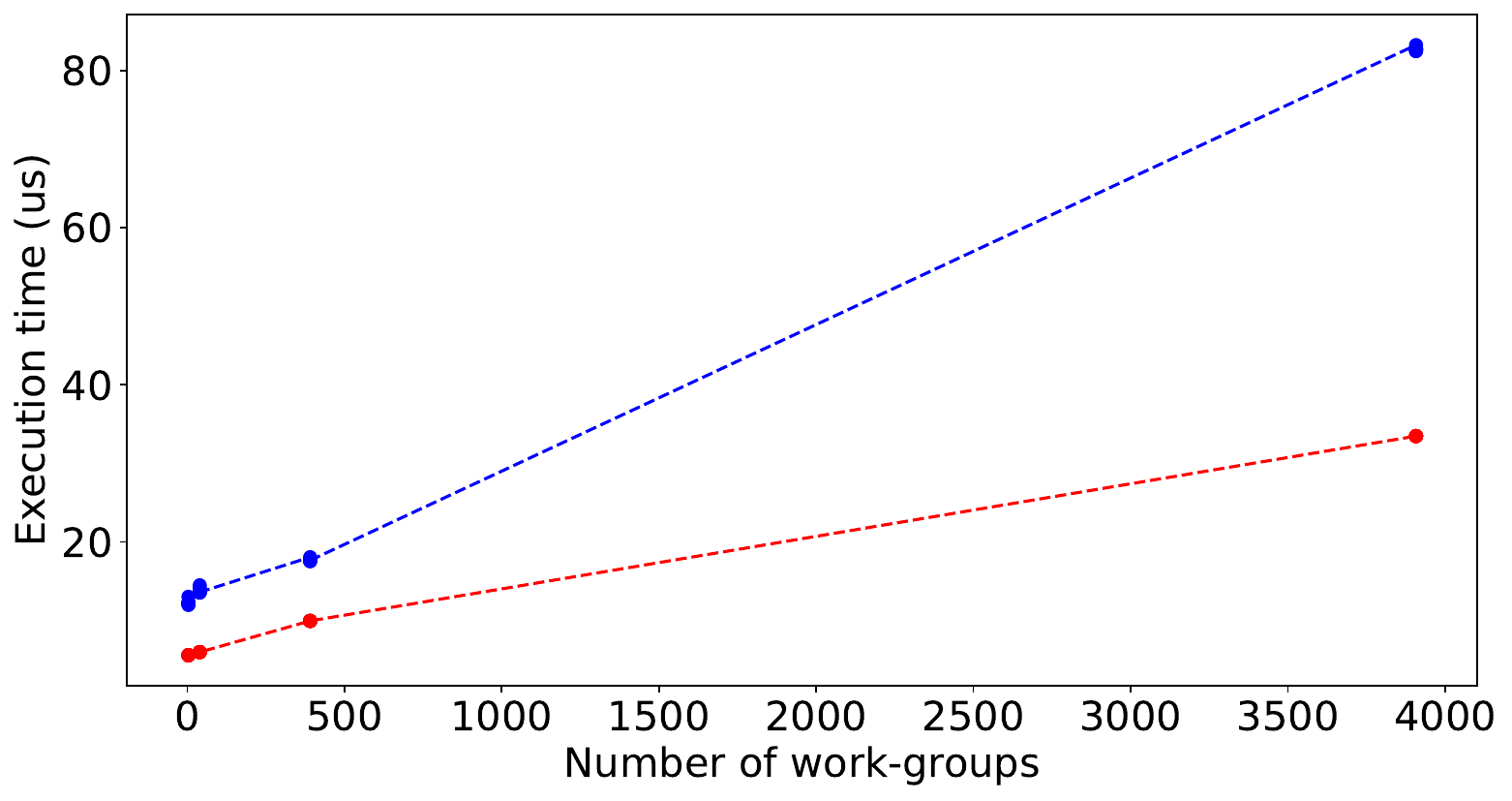}
        \subcaption{SHOC-BFS\_kernel\_warp}
        \label{fig:benchmark-ker1}
    \end{subfigure}
    \hfill
    \begin{subfigure}[b]{0.3\textwidth}
        \includegraphics[width=\linewidth]{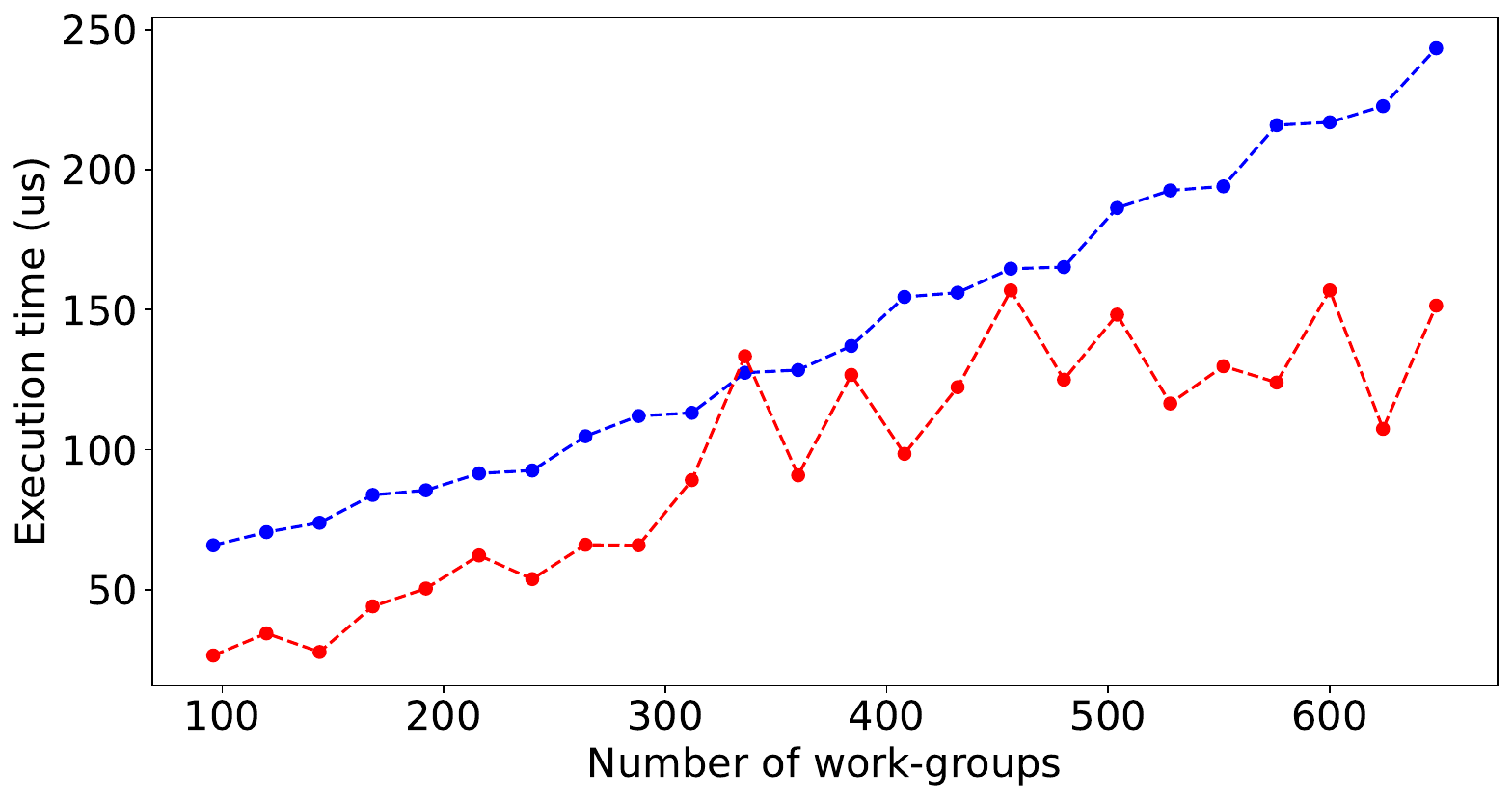}
        \subcaption{SHOC-compute\_lj\_force}
        \label{fig:benchmark-ker2}
    \end{subfigure}
    \hfill
    \begin{subfigure}[b]{0.3\textwidth}
        \includegraphics[width=\linewidth]{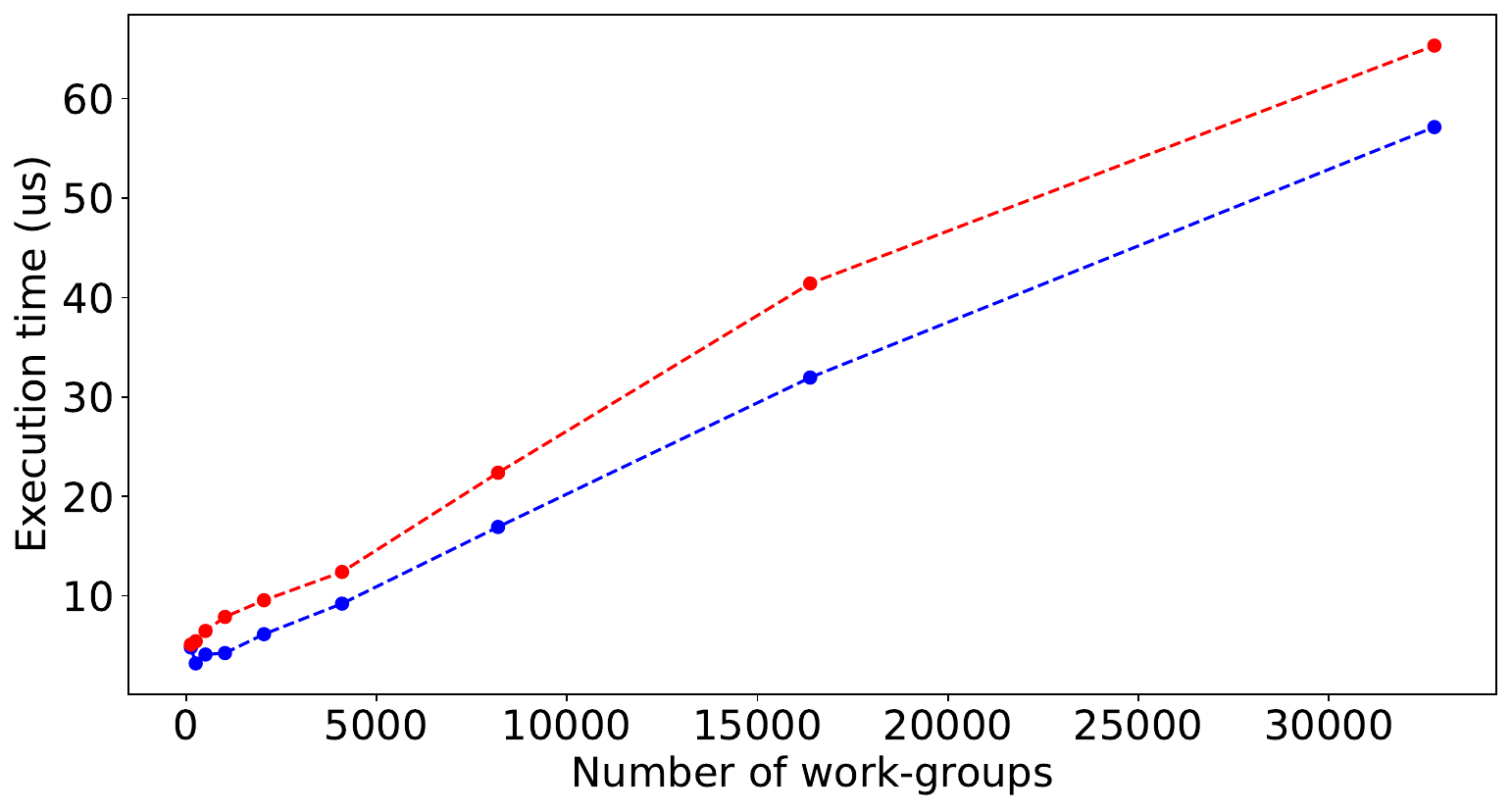}
        \subcaption{SHOC-triad}
        \label{benchmark-ker3}
    \end{subfigure}
    \medskip
    \begin{subfigure}[b]{0.3\textwidth}
        \includegraphics[width=\linewidth]{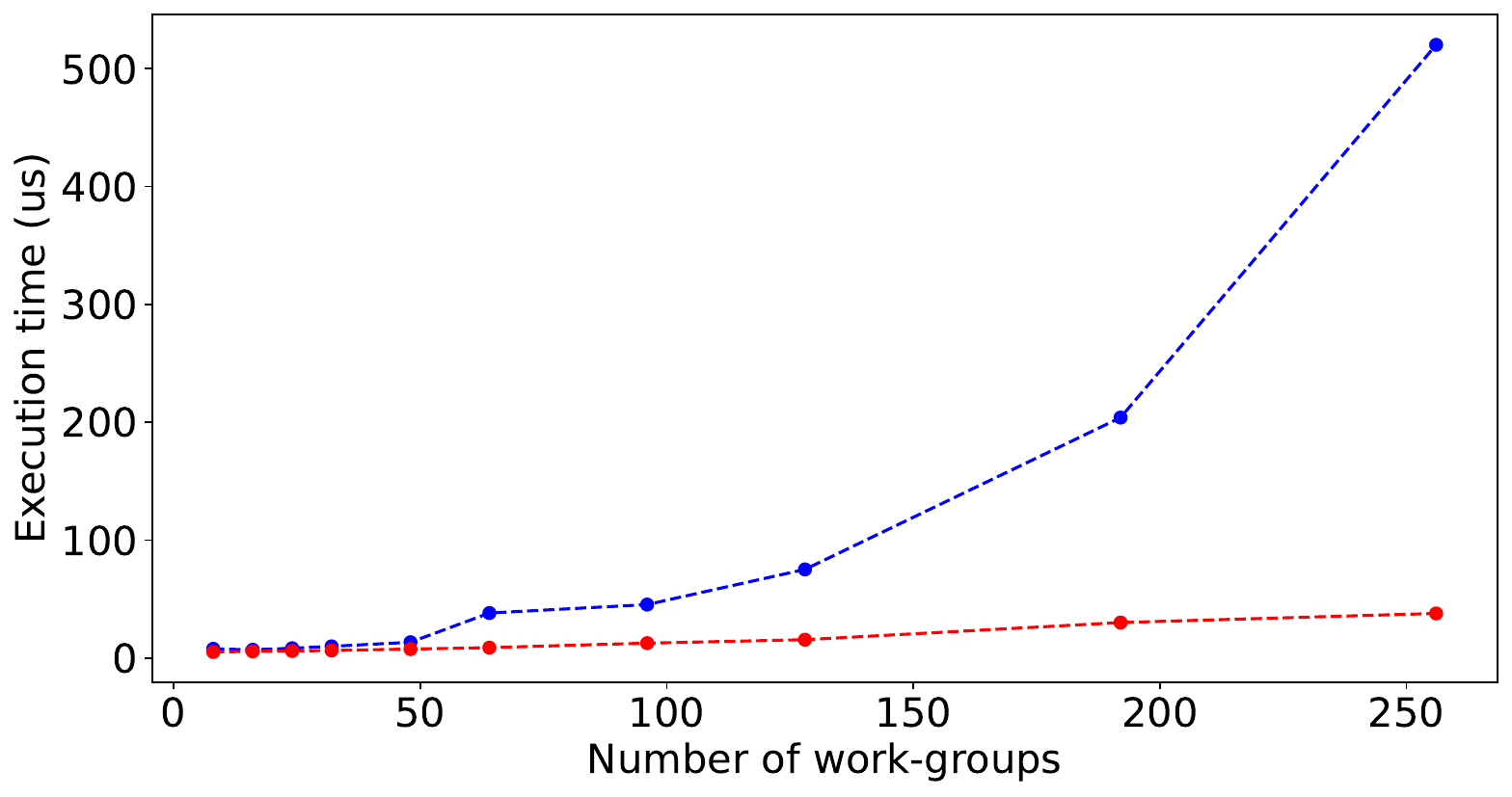}
        \subcaption{SHOC-spmv\_csr\_scalar\_kernel}
        \label{fig:benchmark-ker7}
    \end{subfigure}
    \hfill
    \begin{subfigure}[b]{0.3\textwidth}
        \includegraphics[width=\linewidth]{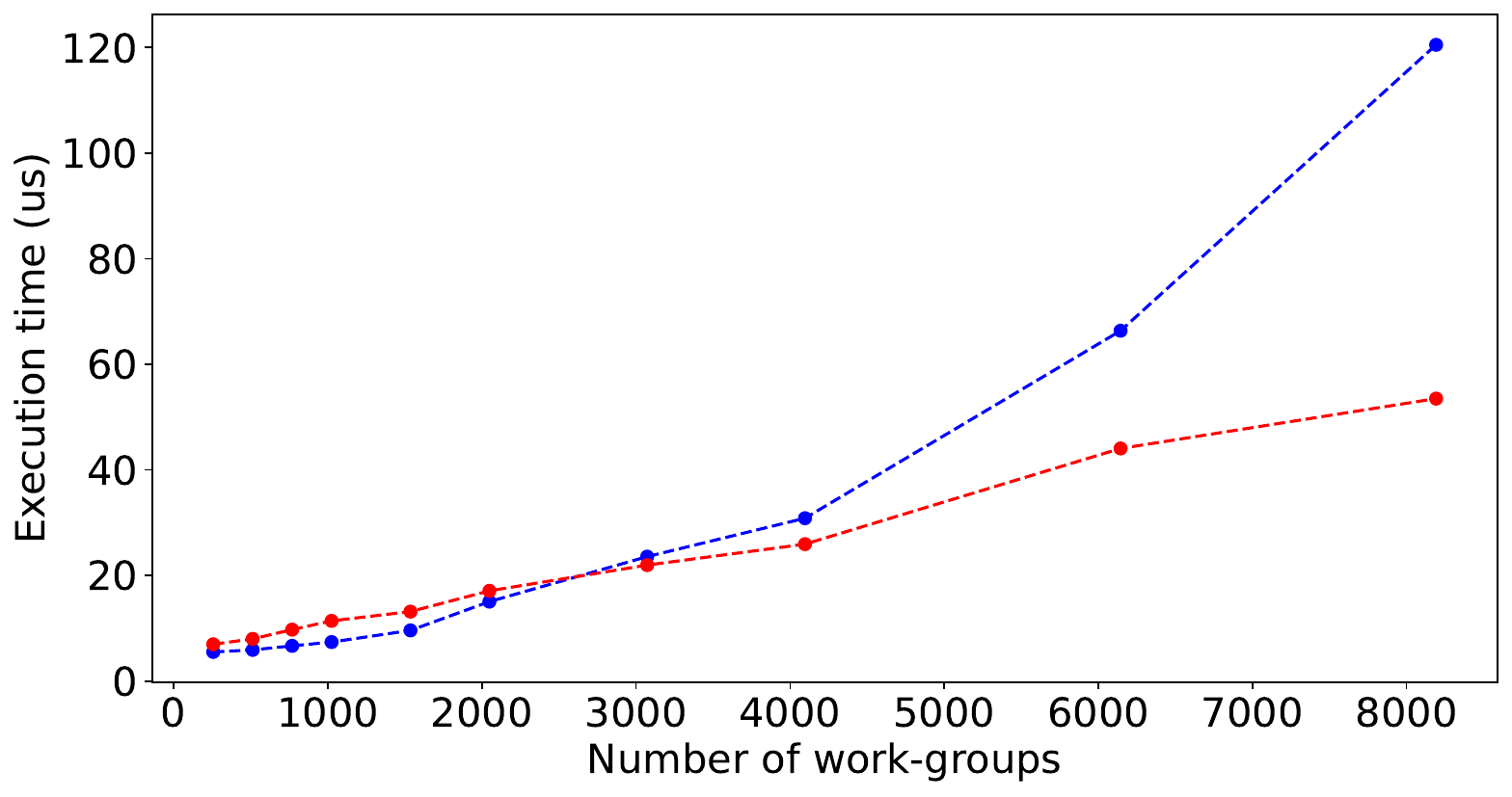}
        \subcaption{SHOC-spmv\_csr\_vector\_kernel}
        \label{fig:benchmark-ker8}
    \end{subfigure}
    \hfill
    \begin{subfigure}[b]{0.3\textwidth}
        \includegraphics[width=\linewidth]{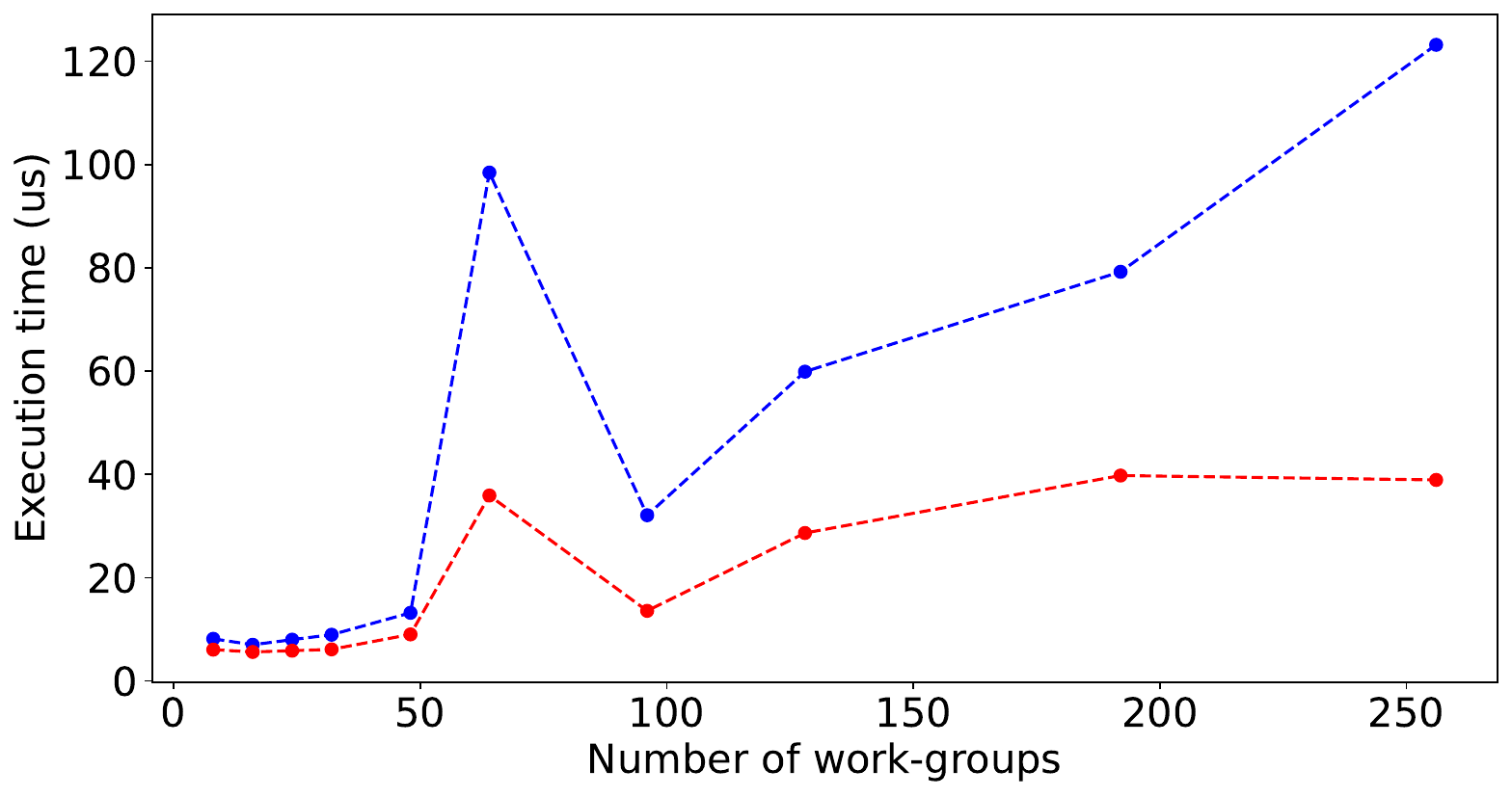}
        \subcaption{SHOC-spmv\_ellpackr\_kernel}
        \label{fig:benchmark-ker9}
    \end{subfigure}
    \medskip
    \begin{subfigure}[b]{0.3\textwidth}
        \includegraphics[width=\linewidth]{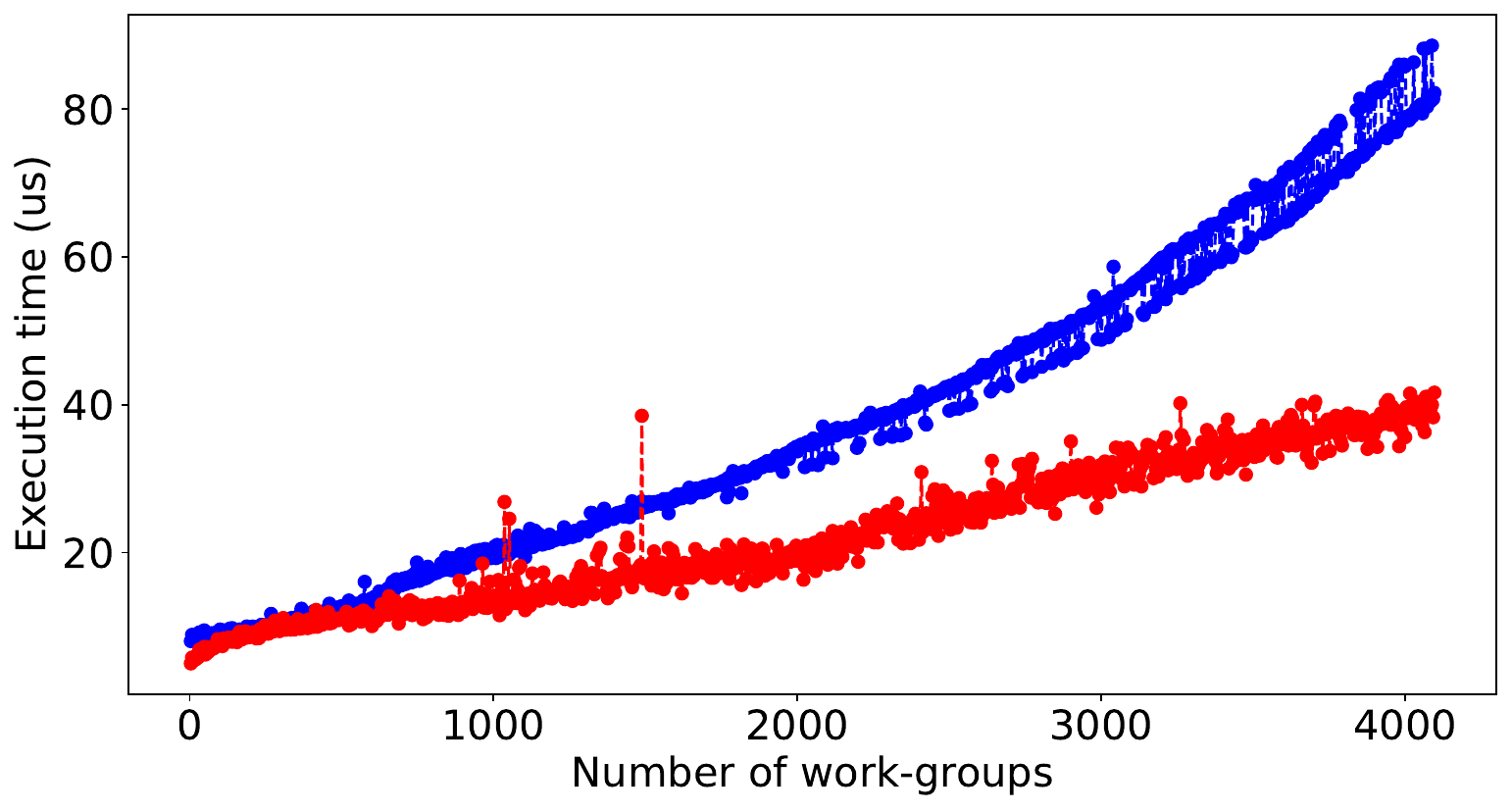}
        \subcaption{Rodinia-BFS\_1}
        \label{fig:benchmark-ker10}
    \end{subfigure}
    \hfill
    \begin{subfigure}[b]{0.3\textwidth}
        \includegraphics[width=\linewidth]{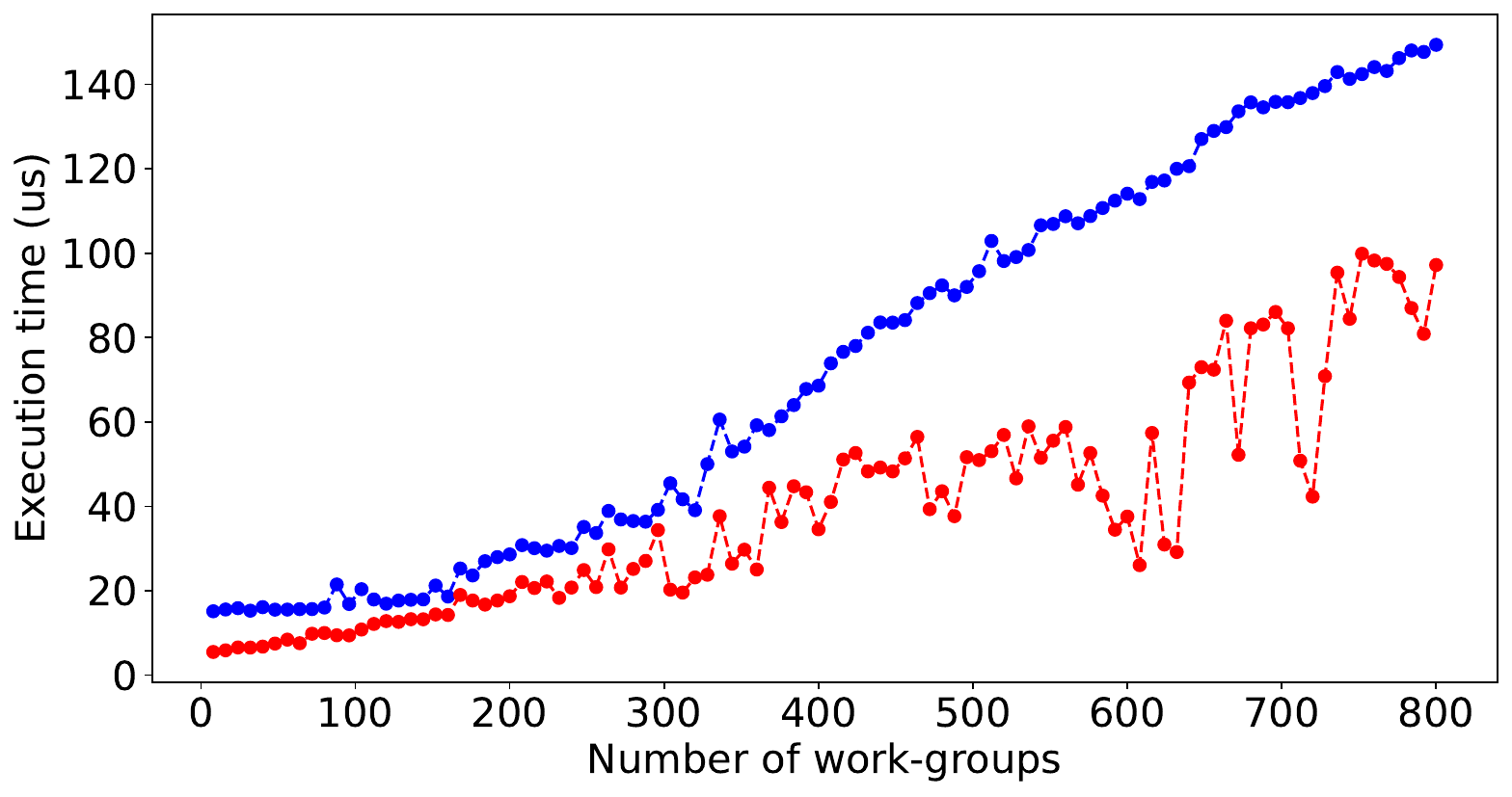}
        \subcaption{Rodinia-kmeans\_kernel\_c}
        \label{fig:benchmark-ker11}
    \end{subfigure}
    \hfill
    \begin{subfigure}[b]{0.3\textwidth}
        \includegraphics[width=\linewidth]{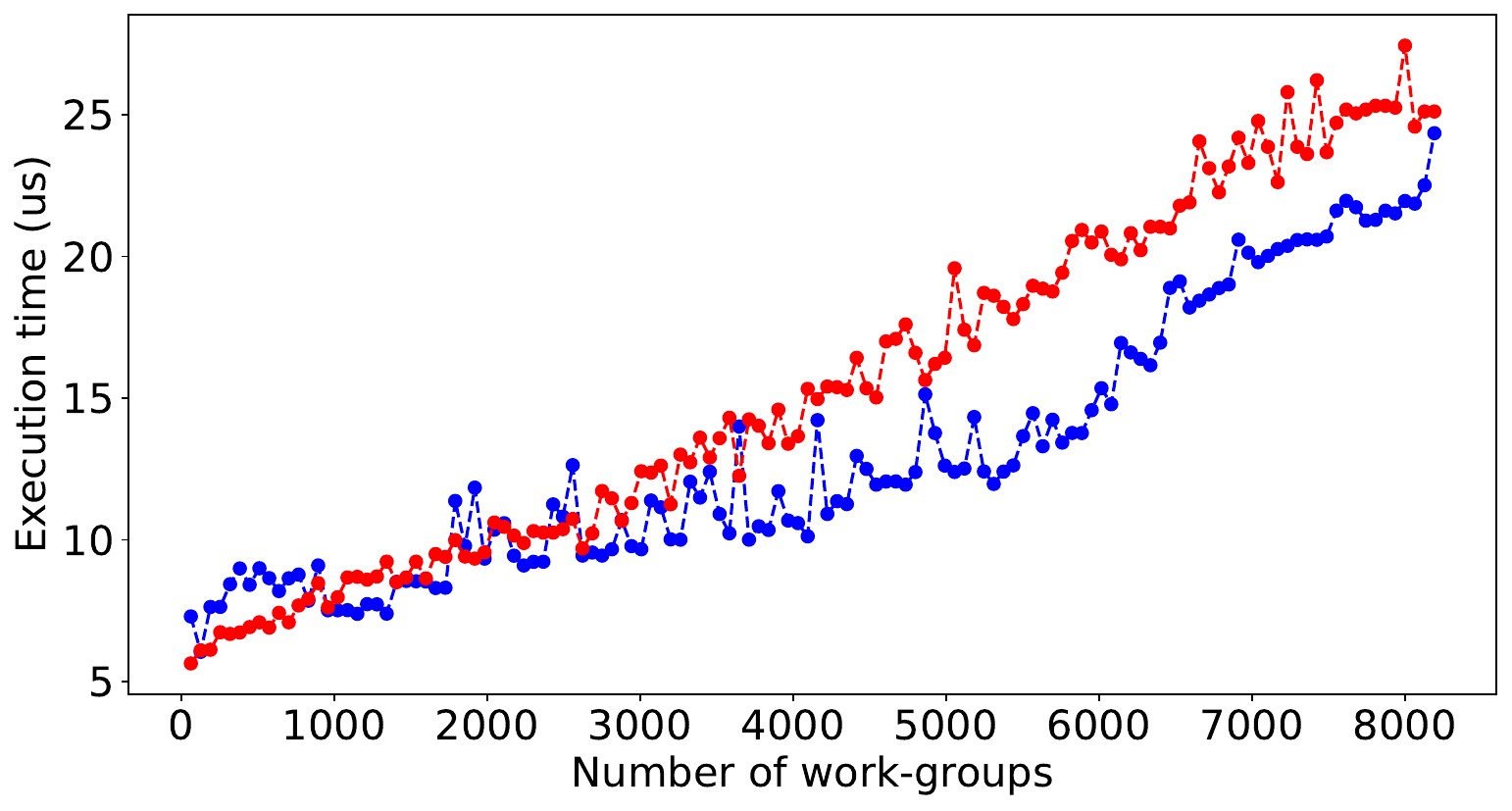}
        \subcaption{Rodinia-NearestNeighbor}
        \label{fig:benchmark-ker12}
    \end{subfigure}
    \medskip
    \begin{subfigure}[b]{0.3\textwidth}
        \includegraphics[width=\linewidth]{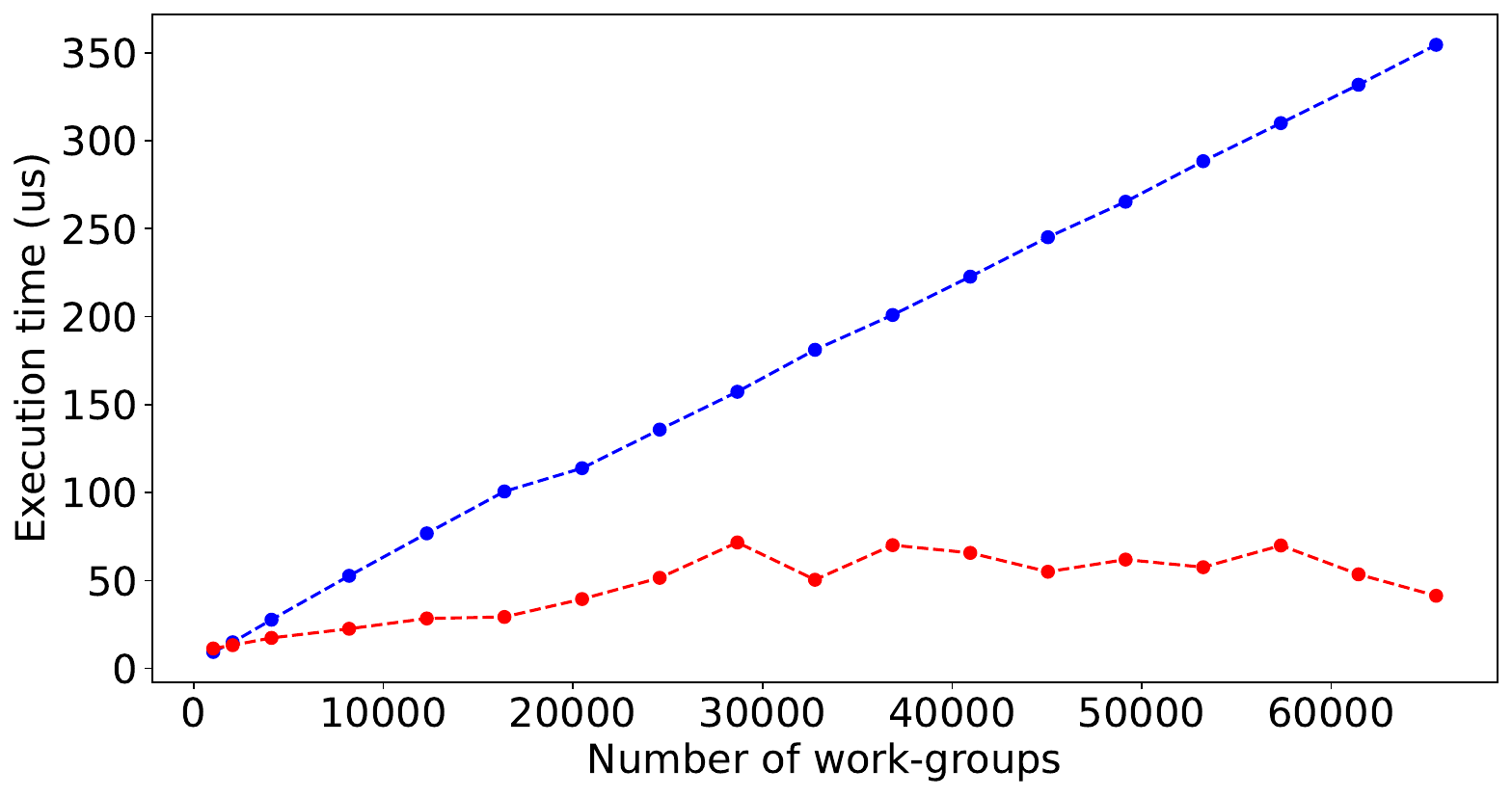}
        \subcaption{SHOC-bottom\_scan}
        \label{fig:benchmark-ker4}
    \end{subfigure}
    \hfill
    \begin{subfigure}[b]{0.3\textwidth}
        \includegraphics[width=\linewidth]{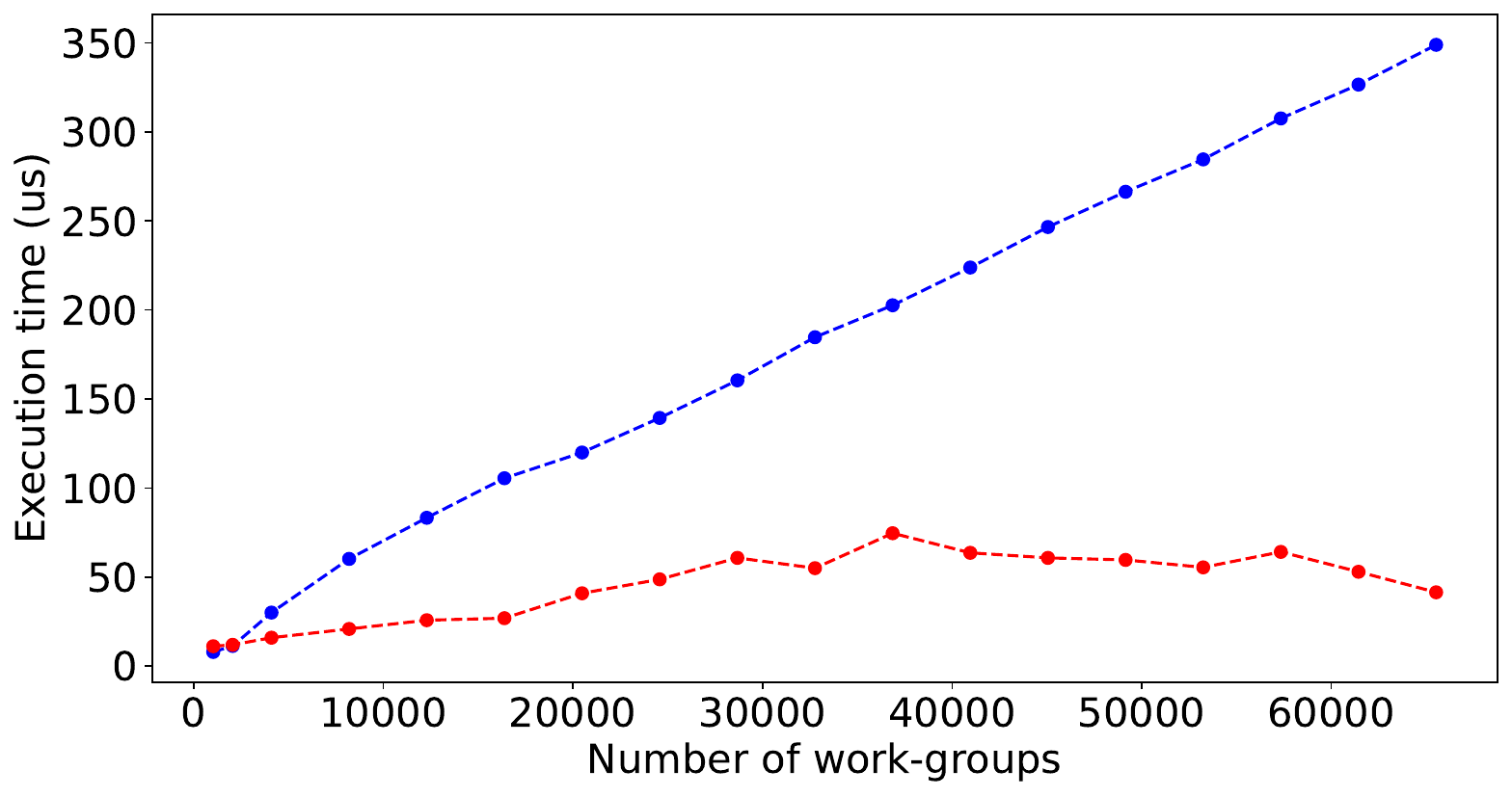}
        \subcaption{SHOC-reduce}
        \label{fig:benchmark-ker5}
    \end{subfigure}
    \hfill
    \begin{subfigure}[b]{0.3\textwidth}
        \includegraphics[width=\linewidth]{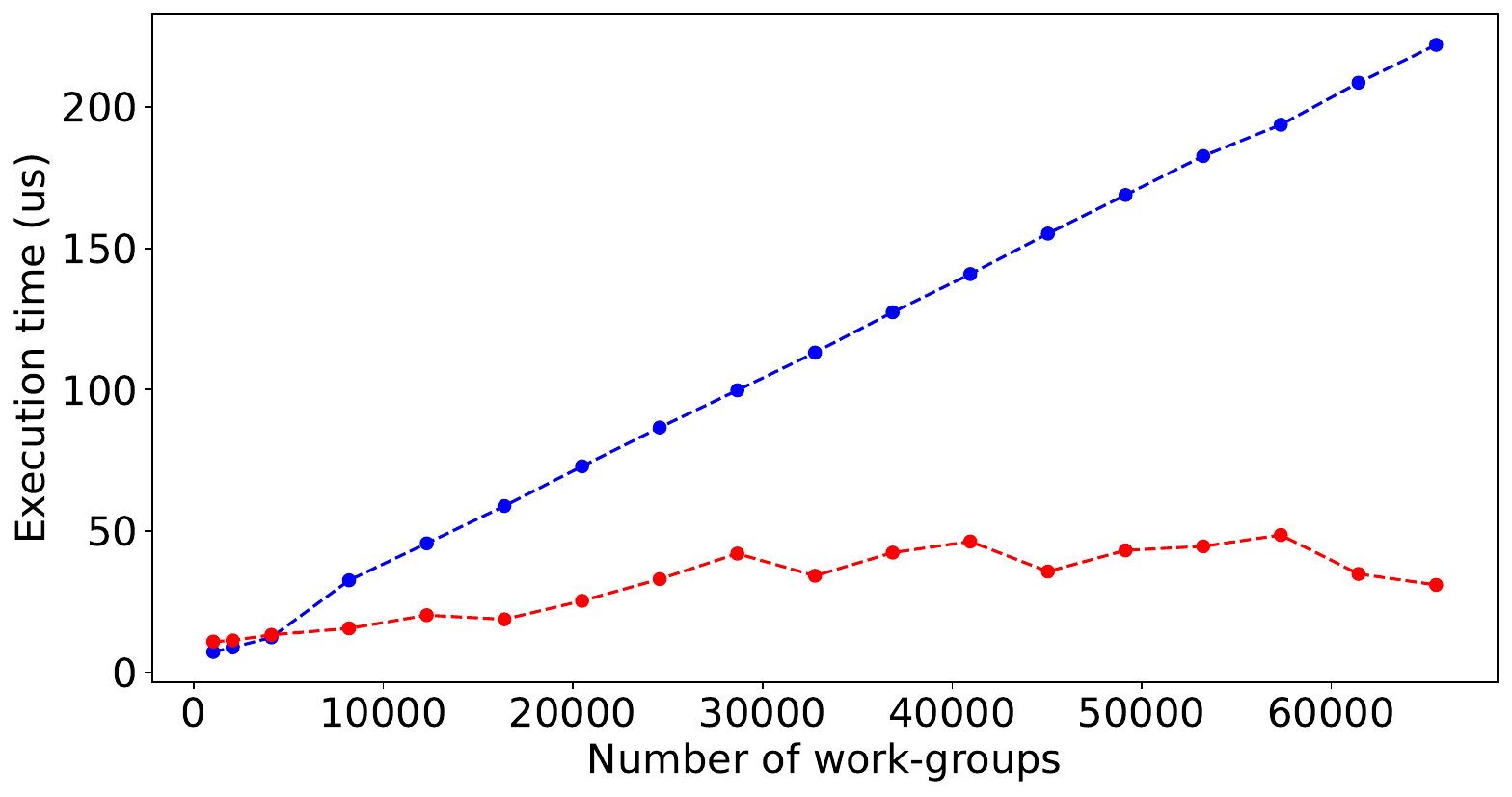}
        \subcaption{SHOC-reduce}
        \label{fig:benchmark-ker6}
    \end{subfigure}
    \caption{Result of LLMPerf on real benchmark kernels by data (input) size. Red points and blue points are prediction time and its corresponding target time.}
    \label{fig:eval-real-benchmark}
\end{figure*}
\subsection{Experiments on large-scale datasets}
\label{sec:evaluation-experiment-large-dataset}
\subsubsection{Experiment settings}

\textbf{Dataset Setting}: LLMPerf is trained on all three datasets.
We split each dataset into training and validation sets with a ratio of $9:1$. We ensure that no kernel appears in both the training and validation sets.
Any prompts exceeding the model's context length are discarded. 
For the target values, execution times are converted to microseconds.
Our datasets are described in Table~\ref{tab:dataset-info}.

\textbf{Model Setting}: We use the \texttt{Multi} variant of the CodeGen model, which is pre-trained on a combination of natural language and code corpora. 
To study the effect of model capacity, we use two model sizes: $350M$ (LLMPerf-350M) and $2B$ (LLMPerf-2B) parameters. The maximum number of tokens (maximum context length) that the model can process is $2048$.

\textbf{Training Setting}: 
The AdamW optimizer \cite{loshchilov2017decoupled} is used with a linear learning rate scheduler, $1000$ warmup steps, a learning rate of $10^{-6}$, $\beta_1 = 0.9, \beta_2 = 0.999, \epsilon = 10^{-8}$, and a weight decay rate of $0.01$. 
Early stopping \cite{morgan1989generalization} is employed to terminate the training process if the loss does not decrease after $15$ epochs.

\textbf{Evaluation metric}: We use Mean Absolute Percentage Error (MAPE) to evaluate the performance of LLMPerf in predicting one particular kernel execution time. MAPE measures the average absolute percentage deviation between the predicted and true (measured) execution times. It is calculated as:
\begin{equation}
    \text{MAPE} = \frac{100\%}{N} \sum_{i=1}^{N} \left\lvert \frac{\hat{y}_i-y_i}{y_i} \right\rvert
\end{equation}
where $N$ is the total number of samples, $\hat{y}_i$ is the predicted execution time, and $y_i$ is the true execution time for sample $i$.

\subsubsection{Experiments result}
Table~\ref{tab:eval-large-dataset} illustrates our results of LLMPerf-350M and LLMPerf-2B on $200K$, $230K$, and $400K$ datasets. After training, we add an additional evaluation of each model on the validation set of the $400K$ dataset to consistently evaluate the models' quality on the most general and unseen dataset among the three validation sets.
LLMPerf-2B-400K demonstrates superior performance on the $400K$ validation set, achieving a MAPE of $24.25\%$, indicating strong generalization and accuracy when trained on the highest quality dataset. In comparison, LLMPerf-2B-200K and LLMPerf-2B-250K exhibit higher MAPEs of $34.2\%$ and $23.61\%$, respectively, though they still outperform their corresponding LLMPerf-350M models in both training and validation sets. This suggests that the LLMPerf-2B models have a greater capacity for learning and generalizing compared to the LLMPerf-350M models.

LLMPerf-350M consistently shows higher validation MAPEs, with $80.53\%$ for LLMPerf-350M-200K, $51.48\%$ for LLMPerf-350M-250K, and $43.77\%$ for LLMPerf-350M-400K, primarily due to its smaller size and limited learning capacity. LLMPerf-350M-400K, despite a low training MAPE of $1.86\%$, suffers from significant overfitting with a validation MAPE of $43.65\%$. The dataset generation techniques also impact model performance: LLMPerf-2B-200K, which had the highest validation MAPE of $64.9\%$, indicates a lack of diversity in performance data, while LLMPerf-2B-250K’s lower MAPE of $23.61\%$ benefits from improved input argument generation. LLMPerf-2B-400K achieves the best result by effectively utilizing a comprehensive dataset, thus overcoming previous limitations and achieving a validation MAPE of $24.25\%$.

\subsection{Experiments on public OpenCL benchmark}
\label{sec:evaluation-experiment-public-benchmark}
Due to the generation technique, in our training dataset, most of our samples employ the input size - global size correlation, where input size equals to global size, only a small amount of them release this correlation. Therefore, we want to assert the model's performance with diverse performance patterns to see its generability. Here, we use the SHOC \cite{danalis2010scalable} and Rodinia \cite{che2009rodinia} benchmark suite to evaluate our best model (LLMPerf-2B-400K). Note that we do not use these benchmarks in our training process.

Usually, the program input size options for each kernel provided by its benchmark are limited, so we vary the input sizes with the same techniques as those used in original benchmarks to further observe the trend of execution time. 
For each input size, we randomly create data, measure the kernel execution time $N$ times (usually $N=100$) and take the average as target execution time. 
We select kernels in these two benchmarks that satisfy our assumptions (are $1D$ kernels and do not exceed the context length of the model) and record their execution times.
\begin{table}[htbp]
    \centering
    \caption{MAPE of LLMPerf-2B-400K on Real Benchmarks}
    \label{tab:benchmark-mape}
    \begin{tabular}{|l|c|}
    \hline
    \textbf{Benchmark-Program (Kernel)} & \textbf{MAPE (\%)} \\
    \hline
    SHOC-BFS (BFS\_kernel\_warp) & 53.62 \\
    SHOC-MD (compute\_lj\_force) & 34.02 \\
    SHOC-Reduction (reduce) & 62.44 \\
    SHOC-Scan (bottom\_scan) & 63.0 \\
    SHOC-Scan (reduce) & 65.85 \\
    SHOC-SPMV (spmv\_csr\_scalar\_kernel) & 56.76 \\
    SHOC-SPMV (spmv\_csr\_vector\_kernel) & 32.51 \\
    SHOC-SPMV (spmv\_ellpackr\_kernel) & 42.76 \\
    SHOC-Triad (triad) & 42.8 \\
    Rodinia-BFS (BFS\_1) & 35.43 \\
    Rodinia-Kmean (kmeans\_kernel\_c) & 43.21 \\
    Rodinia-NN (NearestNeighbor) & 20.93 \\
    \hline
    \textbf{Average} & \textbf{46.11} \\
    \hline
    \end{tabular}
\end{table}
We use LLMPerf-2B-400K model to evaluate all $12$ satisfied kernels. The detailed MAPEs for each kernel are showed in Table~\ref{tab:benchmark-mape}. 
With an average MAPE of $46.11\%$, the model demonstrates reasonably good overall accuracy in predicting kernel execution times only using the kernel source codes. 
The model excels at kernels like \verb|NN| ($20.93\%$) and \verb|MD| ($34.02\%$). 
However, it struggles with kernels such as two \verb|reduce| kernels ($65.85\%$ and $62.44\%$), \verb|bottom_scan|  ($63.0\%$),  \verb|bfs| kernel ($52.62\%$) and \verb|spmv_csr_scalar_kernel| ($53.62\%$).

While MAPE provides a quantitative measure, visualizing predicted and target execution times offers deeper insights into the model's performance. 
Figures 1-9, except for \verb|spmv_csr_scalar_kernel|, roughly predict the relative magnitude between data sizes, albeit slightly lower or higher than the target. 
For kernels with simple performance patterns (no loops or branching) like \verb|triad| or \verb|NearestNeighbor|, the model performs exceptionally well.
Notably, LLMPerf still performs reasonably on kernels with complex patterns, such as loops in \verb|BFS_kernel_warp|, \verb|compute_lj_force|, \verb|spmv_ellpackr_kernel|, \verb|kmeans_kernel_c|, and barriers (thread synchronization) in \verb|spmv_csr_vector_kernel|. 
Even for kernels such as \verb|spmv_ellpackr_kernel|, \verb|spmv_csr_vector_kernel|, and \verb|BFS_1|, where performance depends on array element values (violating our assumption and excluding this information from the prompt), the model demonstrates strong generalization despite limited information provided. This indicates that LLMPerf is robust in understanding kernel performance when sufficient data is available (such as correlations between input size and global size), effectively handling a range of patterns from simple to complex, even with partial information.

However, in cases where a kernel does not have the input size-global size correlation, such as \verb|spmv_csr_scalar_kernel|, it cannot capture the performance pattern despite sharing similarities with \verb|spmv_ellpackr_kernel| and \verb|spmv_csr_vector_kernel|. 
Additionally, when only data size changes, while global size remains fixed, as in \verb|Scan-bottom_scan| (Figure~\ref{fig:benchmark-ker4}), \verb|Scan-reduce| (Figure~\ref{fig:benchmark-ker5}), and \verb|Reduction-reduce| (Figure~\ref{fig:benchmark-ker6}), the model lags in capturing the relationship between input size and performance. 
This limitation likely arises from the relatively small number of varied input size samples ($25K$) compared to the cases where global size and input size are equal ($365K$) in the training set. 
According to the result with input size-global size correlation kernels, we believe increasing the varied input size samples using memory analysis-based input argument generation could potentially overcome this limitation, which we leave for future work.

\section{Conclusion}
\label{sec:conclusion}
In this paper, we introduce the first LLM model that can predict the performance of OpenCL kernels solely based on their source codes.
We also introduce the first large-scale OpenCL performance dataset that we will release publicly.
We introduce techniques using compiler-based memory analysis and statistical analysis to enrich the performance dataset.
We then introduce LLMPerf and evaluate our proposed model on various scenarios.
Although the results that we achieve are humble, we believe a large scale exploration will lead to much better findings, which will be conducted in our future research.
We believe this is the first step to construct a meaningful relationship between NLP models and the system performance.

\section*{Acknowledgement}
We would like to express our sincere gratitude to Moreh Company for providing the computing resources essential for this project. 

\bibliographystyle{IEEEtran}
\bibliography{references}

\end{document}